\documentclass[prl,twocolumn,showpacs,floatfix,aps]{revtex4}
\usepackage{amssymb}
\usepackage{amsmath}
\usepackage{graphicx,bm}
\usepackage{color}
\usepackage{color}

\begin{document}

\title{\textbf{Fluctuation conductivity and possible pseudogap state in FeAs-based  superconductor $EuFeAsO_{0.85}F_{0.15}$}}

\author{A.\,L.\,Solovjov$^{1,2}$, L.\,V.\,Omelchenko$^1$, A.\,V.\,Terekhov $^{1,2}$, E.\,P.\,Khlybov$^{2,3}$
and K.\,Rogacki$^2$}
%\footnote{{Electronic address: solovjov@ilt.kharkov.ua}}
\email{k.rogacki@int.pan.wroc.pl}
\affiliation{$^1$B.\,I.\,Verkin Institute for Low Temperature Physics and Engineering of National Academy of Science of Ukraine, 47 Lenin ave., 61103 Kharkov, Ukraine\\
$^2$International Laboratory of High Magnetic
Fields and Low Temperatures, 95 Gajowicka Str., 53-421, Wroclaw, Poland\\
$^3$L.F. Vereshchagin Institute for High-Pressure Physics, RAS, Troitsk 142190, Russia}

\begin{abstract}
The study of excess conductivity $\sigma'(T)$ in the textured polycrystalline FeAs-based superconductor $EuFeAsO_{0.85}F_{0.15}$ ($T_c = 11\, K$) prepared by the solid state synthesis is reported.
The $\sigma'(T)$ analysis has been performed within the Local Pair (LP) model based on the assumption of the LP formation in cuprate high-$T_c$ superconductors (cuprates) below the pseudogap (PG) temperature $T^*\gg T_c$.
Similarly to the cuprates, near $T_c$ $\sigma'(T)$ is adequately described by the 3D term of the Aslamasov-Larkin (AL) theory
but the range of the 3D-AL fluctuations, $\Delta T_{3D}$, is relatively short.
Above the crossover temperature $T_0 \approx 11.7\,K$ $\sigma'(T)$ is described by the 2D Maki-Thompson (MT) fluctuation term of the Hikami-Larkin (HL) theory.
But enhanced 2D-MT fluctuation contribution, being typical for the magnetic superconductors, is observed.
Within the LP model approach the PG parameter, $\Delta^*(T)$, was determined for the first time.
$\Delta^*(T)$ shows the narrow maximum at $T_s \approx 160\,K$ followed by the linear drop down to $T_{SDW}=T_{NFe}\approx 133\,K$.
Both small $\Delta T_{3D}$ and enhanced $\sigma'(T)$, including linear $\Delta^*(T)$ drop, are considered to be the evidence of the enhanced magnetic interaction in $EuFeAsO_{0.85}F_{0.15}$.
Importantly, the slop of the linear $\Delta^*(T)$  and its length are found to be the same as observed for $SmFeAsO_{0.85}$.
The results suggest both the similarity of magnetic interaction processes in different Fe-pnictides and
applicability of the LP model approach to the $\sigma'(T)$ analysis even in magnetic superconductors.
\end{abstract}

\maketitle

$\bf{1.\, INTRODUCTION}$\\

The discovery of high-Tc superconductivity in
FeAs-based compounds (Fe-pnictides or FePn's) \cite{Kam} has stimulated a great burst of research activity (e.g. see \cite{Sad,Jo,Pag,St} and references therein).
Following the discovery in LaFeAs(O,F) with $T_c$ = 26 K \cite{Kam}, superconductivity
was found in many materials with related crystal structures, that commonly possess iron-pnictide or iron-chalcogenide layers.
Actually the various members of the iron containing FePn's can be divided into  three main family of materials, which show superconducting (SC) transition upon substitution by a dopant or upon applying external pressure.
They are, (i) the quaternary 1111 compounds, RFeAsO, where R represents a lanthanide such as La, Ce, Sm, Eu etc. \cite{Kam,Tak,Che1,Che2} with transition temperatures as high as 56\,K in $SmFeAsO_{1-x}F_x$;
(ii) the ternary $AFe_2As_2$ (A = Ca, Sr, Ba, Eu) \cite{Ro,Jee1,Sas,Wu} systems, also known as 122 systems that exhibit superconductivity up to 38\,K;
and (iii) the binary chalcogenide 11 systems (e.g. FeSe) with superconducting transition temperatures up to 14\,K \cite{Hsu}.
The common feature for all families is  a structural transition from a
tetragonal to an orthorhombic phase at $T_s = (150 - 190)\,K$ which is closely related
to the formation of a spin-density-wave (SDW) type magnetic instability at T = $T_{SDW}$ due to antiferromagnetic (AF) ordering of the Fe spins \cite{Sad}.
For "1111" systems $T_{SDW} < T_S$ \cite{Sad,S1} whereas for "122" compound, e.g for $EuFe_2As_2$ \cite{Tok},
$T_{SDW} \approx T_s$.
Apparently, the superconductivity emerges from the FeAs or FeSe layers which are the building blocks of the corresponding quasi two-dimensional crystal structures suggesting the analogy to the cuprate-based high-$T_c$ superconductors (HTS's).
Like in the cuprates and heavy fermion metals, superconductivity of the iron-based compounds has a direct relation to magnetism.
The maximal $T_c$ is found in the vicinity of the extrapolated point where SDW order of the Fe 3d magnetic moment is suppressed by doping or pressure.

\indent However, also like in the cuprates, up to now the physical nature of the superconducting pairing mechanism in the new FeAs-based HTS's remains uncertain \cite{Av}.
There is a growing evidence that, in addition to the common electron-phonon (e-ph) interaction \cite{Ma}, it is of presumably magnetic type \cite{St,Av}, and all members of the iron arsenide $RFeAsO_{1-x}F_x$ family are characterized by the long-range (non-local) magnetic correlations \cite{M}.
It is well known that upon electron or hole doping with F substitution at the
O site \cite{Kam,Rig,Hu} or with oxygen vacancies \cite{Now,Ren}, all properties of the parent RFeAsO compounds drastically change and evident AF order has to disappear \cite{Sad}.
However, recent results \cite{Dr,San,Che3,Zhu,Fer} point toward an important role of the low-energy spin magnetic fluctuations \cite{EM}.
They emerge on doping away from the parent AF state which is of a SDW type \cite{Dr,San,Fer} as mentioned above.
Thus, below $T_S$ the AF fluctuations, being likely of spin wave type, are believed to noticeably affect the properties of doped $RFeAsO_{1-x}F_x$ systems \cite{M,Dr,San}.
As shown by many studies \cite{Dr,San,Che3,Zhu}, the static magnetism persists well
into the SC regime of FePn's.
As a result, rather peculiar normal state behavior of the doped systems upon T diminution is
expected in this case \cite{Rig,Che3,Zhu}.
Besides, it was recently shown theoretically that antiferromagnetism and superconductivity can coexist in these materials only if Cooper pairs form an unconventional, sign-changing state \cite{M,Zhu,Fer}.

\indent The correlation between the SDW and SC order is a central topic in the current research on the FeAs-based high-$T_c$ superconductors.
However, the clear nature of the complex interplay between the magnetism and the
superconductivity in FeAs-based HTS's is still rather controversial.
As a result, rather complicated phase diagrams for different FePn's \cite{Che3,Zhu,Fer,Hua} and especially for $SmFeAsO_{1-x}F_x$ \cite{Rig,Kam2} are reported.
For all these HTS's rather wide temperature region is found in which superconductivity coexists with SDW regime.

\indent In this paper we focus on the study of the fluctuation conductivity (FLC) and possible pseudogap (PG) in $EuFeAsO_{0.85}F_{15}$.
Somewhat surprisingly, among the quaternary "1111" compounds $EuFeAsO_{1-x}F_x$ is not enough studied.
It is likely due to the largest atomic radius of Eu, $r_{at} \approx 2.1 \AA$, resulting in relatively low
$T_c \approx 11 \,K$ and $H_{c2} \approx 14 \, T$ at $0.7\,T_c$ \cite{Dm}.
In this case just the ternary Eu-based "122" compounds such as $Eu_{0.5}K_{0.5}Fe_2As_2$ ($T_c$ = 32\,K) \cite{Jee1},  $EuFe_2(As_{1-x}P_x)_2$
($T_c \approx$ 28\,K) \cite{Jee2} and $EuFe_{2-x}Co_xAs_2$ ($T_c \approx$ 21\,K) (see \cite{Kru} and references therein) were widely studied.
Special attention was devoted to $EuFe_2As_2$ because it is the only rare-earth based member of the "122" family.
Besides, in contrast to the $AFe_2As_2$ (A=Ca, Sr, Ba) compounds where only the iron possesses a magnetic moment, in $EuFe_2As_2$ a large additional magnetic moment of about $7\mu_B$ is carried by Eu which is in the 2+ state.
As a result, it exhibits a combined transition of structural and SDW order of Fe magnetic moments at the highest reported  $T_S=T_{SDW} \approx 190\,K$ in the FePn's and subsequently Eu 4f moments order below $T_{NEu}\approx 20$\,K into a canted AF state \cite{Jee1,Ren}.
Thus, in this system it seems to be possible to study the interplay between the localized $Eu^{2+}$ moments and the itinerant magnetism of the FeAs layers along with its influence on superconductivity under application of the
hydrostatic pressure or doping.
Besides, it was found that the AF ground state could easily be switched to a FM state in small in-plane fields of order 1\,T \cite{Tok}.
These observations suggest that the Eu-based systems are close to a FM instability \cite{Tok,Dm}.
Thus, many different properties of the parent as well as of doped $EuFe_2As_2$, from relatively simple resistivity measurements \cite{Jee1} up to angle resolved photoemission spectroscopy (ARPES) studies which revealed the droplet-like Fermi surfaces in the AF phase of $EuFe_2As_2$ \cite{Jon}, were thoroughly analyzed.
At the same time, the properties of $EuFeAsO_{1-x}F_x$ remain somewhat uncertain \cite{Dm,Dm2}.

\indent Moreover, despite of the number of papers devoted to the FeAs-based superconductors, in contrast to cuprates, there is an evident lack of the fluctuation conductivity (FLC) and pseudogap (PG) studies in FePn's \cite{Sad}.
Strictly speaking, apart from our investigation of the FLC and PG in $SmFeAsO_{0.85}$ \cite{S1} we have no information
about the similar experiments performed by another research groups.
As a result, the possibility of a PG state in the FeAs-based HTS's still remains controversial \cite{AK}.
It is well known that the pseudogap is a specific state of matter which is observed in underdoped  cuprates and characterized by reduced density of states (DOS) at the Fermi level at temperatures well above $T_c$ \cite{AK,Deu,Mish}.
For $YBa_2Cu_3O_{7-\delta}$ (YBCO) the noticeable reduction of DOS, i.e pseudogap, was observed below representative PG temperature $T^*\gg T_c$ in the study of the Knight shift measured by NMR \cite{K}.
Recently reduced DOS and PG were directly measured  by ARPES for $BiSCCO$ \cite{Kon}.
Unfortunately, there is no information about such experiments performed on FePn's.
\begin{figure}[b]
%\noindent\centering{
\begin{center}
\includegraphics[width=.48\textwidth]{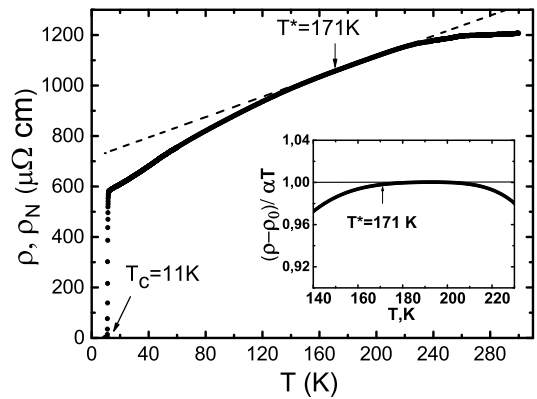}
\caption{Temperature dependence of the in-plane resistivity $\rho$ for $EuFeAsO_{0.85}F_{0.15}$ (dots). Straight dashed line designates extrapolated normal-state resistivity $\rho_N(T)$. Insert: $(\rho-\rho_0)/\alpha T$ vs T which provides the more precise determination of $T^*=171~K$.}
\end{center}
%\label{ResTr2}
\end{figure}

Nevertheless, electron spin resonance (ESR) of $Eu^{2+}$ which successfully probes the local DOS of the conduction electrons in the normal state ($T>T_{SDW}$) have recently been measured on
$EuFe_{2-x}Co_xAs_2$ $(0 \le x \le 0.4)$ and $EuFe_2As_{2-y}P_y$ $(0 \le y \le 0.43)$ iron pnictides \cite{Kru}.
It was shown that substitution of cobalt for iron or phosphorous for arsenic
suppressed gradually the SDW phase and reduced the slope of the linear increase
of the linewidth $\Delta H(T)$ above $T_{SDW}$, due to the Korringa  relaxation,
down to about b = 3\, Oe/K.
This indicates the reduction of the conduction-electron DOS at the Fermi energy on increasing Co or P substitution.
The fact suggests the possibility of the PG state in doped FePn's, at least in the Eu-based compounds.

\indent To clarify the issue, we have analyzed the excess (fluctuation) conductivity derived from the resistivity measurements on $EuFeAsO_{1-x}F_x$.
The analysis has been performed within our local pair (LP) model \cite{S2,Tk}.
The model based on the assumption that in cuprate HTS's the PG is due to formation of the local pairs below $T^*$ \cite{Mish,L,H,Eng}.\\

\indent {\bf 2.\, EXPERIMENT}\\

Textured polycrystalline samples of $EuFeAsO_{0.85}F_{0.15}$ were synthesized by solid state reaction method as described elsewhere \cite{Jee1,Dm}.
Rectangular samples of about 5x1x1\,mm were cut out of the pressed pellets.
A fully computerized setup on the bases of a Physical Properties Measurement System
(Quantum Design PPMS-9T) utilizing the four-point probe technique was used to measure the longitudinal resistivity, $\rho_{xx}(T)$.
Silver epoxy contacts were glued to the extremities of the sample in order to produce a uniform current distribution in the central region where voltage probes in the form of parallel stripes were placed.
Contact resistances below $1\Omega$ were obtained.
Temperature dependence of resistivity $\rho(T)=\rho_{xx}(T)$ for studied $EuFeAsO_{0.85}F_{0.15}$ with $T_c$ = 11\, K is shown in Fig. 1 (dots).

The superconducting transition temperature $T_c$ is determined by extrapolation of the linear part of the resistive transition to $\rho(T_c)=0$ \cite{Lang}.
The comparatively small width of the SC transition rules out significant
variation of the superconducting parameters over the sample volume.
The whole resistivity curve (Fig. 1) is somewhat S-shaped with the feebly marked positive thermally activated buckling characteristic for the slightly doped cuprates \cite{And}.
However, over the temperature range $T^*\approx 171~K$ to $T \approx 210~K$ $\rho(T)$ varies linearly with T at rates $d\rho/dT$ = 2.0 $\mu\,\Omega\,cm\, K^{-1}$.
Above 210 K $\rho(T)$ exhibits a downwards deviation from the linear dependence which is typical for the FePn's \cite{Kam,Sad,S1} (refer to Fig. 1).
The linear dependence can be written as $\rho_N(T) = \alpha T+\rho_0$, where $\rho_N(T)$ is the normal state resistivity extrapolated to low $T$ region \cite{S2,SP} and $\rho_0$ is its intercept with y-axis.
It is evident that $(\rho(T)-\rho_0)/\alpha T$=1 above the PG temperature $T^*$ providing the more precise way of $T^*$ determination \cite{DeM}.
Insert in Fig. 1 demonstrates the result of this approach.\\

\indent {\bf 3.\, RESULTS AND DISCUSSION}

\indent {\bf 3.1.\, Fluctuation conductivity}\\

Below the representative temperature $T^*=(171\pm 1.0)\,K \gg T_c$ (Fig.1) resistivity of $EuFeAsO_{0.85}F_{15}$, $\rho$(T), deviates down from linearity resulting in appearance of the excess conductivity as a difference between measured $\rho(T)$  and extrapolated normal state resistivity $\rho_N(T)$:
\begin{equation}
\sigma '(T) = \sigma(T) - \sigma_N(T)=[1/\rho(T)]-[1/\rho_N(T)],
\label{sigma-t}
\end{equation}
Mentioned above procedure of the normal state resistivity $\rho_N(T)$ determination by the linear dependence is widely used in the literature (see \cite{S2,Lang,DeM,Oh,ND} and references therein) and has been justified theoretically by the Nearly Antiferromagnetic Fermi Liquid (NAFL) model \cite{SP}.

\indent In the case of cuprates, $\sigma '(T)$ is intimately connected with the PG \cite{Vovk,S2,Tk} and is believed to appear due to formation of the local pairs (LP) at
$T\le T^*$ regarded as a pseudogap temperature \cite{Mish,L,H,Eng,Kag}.
As mentioned above, there are several experiments \cite{K,Kon} in which the partial decrease of DOS at the Fermi level, which is just called a PG \cite{AK,S2,L}, was observed in cuprates below $T^*$.
In the case of FePn's the magnetic subsystem is believed to be also taken into account to explain the excess conductivity appearance.
It is especially the case for $EuFeAsO_{0.85}F_{0.15}$ where the iron magnetic moment is added by the large magnetic moment carried by Eu which is believed to be partly in the 2+ state \cite{Dm2}.
Thus, in FePn's the excess conductivity is expected to be due to both LP formation and a specific magnetic interaction which has to somehow govern the LP behavior.
This process, however, is weakly studied.
Unfortunately, except for the mentioned above ESR measurements on Eu-based FePn's \cite{Kru}, there are no direct experiments on measuring temperature dependence of DOS in FePn's.
Thus, the question as for the possibility of a PG state in FePn's also remains uncertain.
Besides, up to now there is no rigorous theory to describe the excess conductivity in the whole temperature range from $T^*$ down to $T_c$ in the HTS's.
That is why we have tried to analyze found $\sigma '(T)$ within our LP model approach paying more attention at the possible difference in revealed results and parameters in comparison with those obtained for YBCO films \cite{S2} and $SmFeAsO_{0.85}$ polycrystals  \cite{S1, S2}.
Here we focus on the analysis of the fluctuation conductivity (FLC) and possible pseudogap (PG) derived from measured excess conductivity within our LP model \cite{S2,Tk}.
Determined from the analysis sample parameters are listed in the Table.
\begin{figure}[b]
%\noindent\centering{
\begin{center}
\includegraphics[width=.50\textwidth]{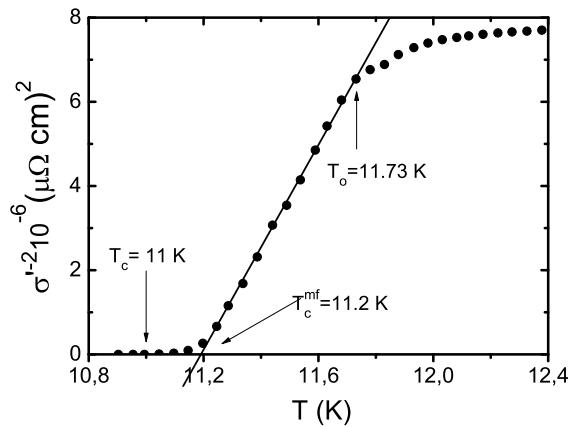}
\caption{Inverted squared excess conductivity $\sigma'^{-2}$ (dots) as a
function of temperature plotted in the temperature range near $T_c$.
The intercept of its linear extrapolation with the x-axis determines $T_c^{mf} =11.2~K$. Solid line is a guidance for eye.}
\end{center}
%\label{ResTr2}
\end{figure}

\indent Our LP model approach consists of several logical steps \cite{S2}.
First, the mean field critical temperature $T_c^{mf}$, which determines the reduced temperature \cite{HL}
\begin{equation}
\varepsilon = (T - T_c^{mf})\,/\,T_c^{mf}
\label{var}
\end{equation}
and is of a primarily importance for the whole analysis, must be defined.
Here $T_c^{mf}>T_c$ is the critical temperature in the mean-field approximation, which separates the FLC region from the region of critical fluctuations or fluctuations of the order parameter $\Delta$ directly near $T_c$ (where $\Delta<k_BT)$ neglected in the Ginzburg-Landau (GL) theory \cite{DeGen}.
As shown by many studies (see \cite{S2,S3} and references therein)
FLC near $T_c$ is always extrapolated by the standard equation of the Aslamasov-Larkin (AL) theory \cite{AL} with the critical exponent $\lambda=-1/2$ (Fig. 3, line 1) which determines the FLC in any 3D system
\begin{equation}
\sigma_{AL3D} '=C_{3D}\frac{e^2}{32\,\hbar\,\xi_c(0)}{\varepsilon^{-1\,/\,2}}.
\label{3D}
\end{equation}
Here $C_{3D}$ is a numerical factor used to fit the data by the theory \cite{S3,Oh}
and $\xi_c(T)$ is a coherence length along the c-axis \cite{HL}.
This means that the conventional 3D FLC is realized in HTS's as $T \to T_c$ \cite{S3,Xie}.
From Eq. (3), one can easily obtain $\sigma'^{-2}\sim (T - T_c^{mf})\,/\,T_c^{mf}$.
Evidently,  $\sigma'^{-2}=0$ when $T = T_c^{mf}$.
This way of $T_c^{mf}$ determination was proposed in Ref. \cite{Beas} and justified by different FLC experiments \cite{S2,Oh,S3}.
Moreover, when $T_c^{mf}$ is properly chosen the data in the 3D fluctuation region near  $T_c$ are always fitted by Eq. (3) \cite{S2}.

\indent Figure 2 displays the $\sigma'^{-2}$ vs T plot for our $EuFeAsO_{0.85}F_{0.15}$ (dots).
The intercept of the extrapolated linear $\sigma'^{-2}$ with T-axis determines $T_c^{mf}= 11.2~K$.
Above the crossover temperature $T_0 \approx 11.73~K$ the data deviates right from the line suggesting the presence of the 2D Maki-Thompson (MT) \cite{Mak,Tho} fluctuation contribution to $\sigma'(T)$ \cite{S3,Xie}.
At the crossover temperature $T_0 \sim \varepsilon_0$ the coherence length $\xi_c(T)=\xi_c(0)\varepsilon^{-1/2}$ is expected to amount to d, which is the distance between conducting layers in HTS's \cite{Lang,Oh,HL}.
This yields
\begin{equation}
\xi_c(0) = d\,\sqrt\varepsilon_0
\end{equation}
and allows to determine $\xi_c(0)$ which is one of the important parameter of the PG analysis.

\indent The excess conductivity $\sigma'$, derived from the resistivity measurements by means of Eq. (1), is plotted in Fig. 3 (dots) as a function of $\varepsilon$  in customary double logarithmic scale.
As expected, above $T_c^{mf}$ and up to $T_0$ = 11.73~K ($\ln \varepsilon_0 \approx -3.04$) $\sigma'$ vs $\epsilon$ is well fitted by the 3D fluctuation term (3) of the AL theory (Fig. 3, solid line 1) with $\xi_c(0)= (2.84 \pm 0.02)\AA$ determined by Eq. (4) (see the Table) and $C_{3D}$ = 0.32.
By analogy with the cuprates, to find $\xi_c(0)$ we make use of d=c, which is the c-axis lattice parameter \cite{S2,Lang}.
Unfortunately, it is not much known about the lattice parameters in $EuFeAsO_{0.85}F_{15}$.
That is why, in contrast to cuprates, where $d\approx 11.7\AA$ \cite{S3}, now we set d=c=13\AA\ being determined for $Eu_{0.5}K_{0.5}Fe_2As_2$ \cite{Jee1}.
\begin{figure}[t]
%\noindent\centering{
\begin{center}
\includegraphics[width=.50\textwidth]{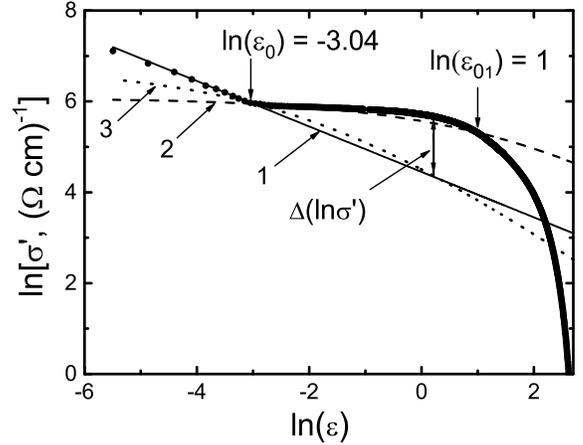}
\caption{$ln\sigma'$ as a function of $ln\varepsilon$ (dots) compared with fluctuation theories: 1- 3D AL; 2 - MT with $d=d_{01}$; 3 - MT with $d=13$\AA. $ln\varepsilon_0=-3.04$ corresponds to the crossover temperature $T_0$ and allows to determine $\xi_c(0)=d\sqrt\varepsilon_0=(2.84 \pm 0.02)\AA$. Accordingly, $ln\varepsilon_{01}=1$ corresponds to the representative temperature $T_{01}$ below which the Josephson interaction between the internal planes has to set in.}
\end{center}
%\label{ResTr2}
\end{figure}

Found $\xi_c(0)= (2.84 \pm 0.02)\AA$ is about 1.7 and 2.0 times of that obtained for the YBCO film ($T_c = 87.4\,K$) \cite{S2} and correspondingly for $SmFeAsO_{0.85}$ polycrystal ($T_c = 55\,K$) \cite{S1}, which are considered to be the reference samples (Table I).
It is not surprising seeing we assume $\xi(0)\sim \hbar v_F/\pi\Delta(0)$ or $\xi(0)\sim 2\hbar v_F/5\pi k_BT_c$ \cite{S4}.
Here we have taken into account the experimental \cite{S5} and theoretical \cite{Z,WCh} fact that $2\Delta(0)/k_BT_c \sim 5$ for YBCO HTS's, which is the sign of the strong superconductivity in contrast with the weak BCS superconductivity, where $2\Delta(0)/k_BT_c = 3.52$.
Thus, the lower $T_c$ the higher both $\xi_c(0)$ and correspondingly the in-plane coherence length $\xi_{ab}(0)$ in agreement with our results.
Simultaneously, the temperature range of the 3D FLC turned out to be rather small: $\Delta T_{3D}\approx 0.5\,K$ (Fig. 2 and 3).
Accordingly, $\Delta T_{3D}\approx 1.8\,K$ and $\Delta T_{3D}\approx 1.5\,K$ are obtained for the reference YBCO film \cite{S2} and $SmFeAsO_{0.85}$ polycrystal \cite{S1}, respectively.
The shortening of the $\Delta T_{3D}$ can likely be considered as a first sign of the enhanced magnetic interaction in $EuFeAsO_{0.85}F_{0.15}$.
The conclusion comes from the fact that the shortest $\Delta T_{3D} \approx 0.16\,K$ was observed for the wholly magnetic superconductor
$Dy_{0.6}Y_{0.4}Rh_{3.85}Ru_{0.15}B_4$ with $T_c = 6.4\,K$ \cite{AT} (see the Table).
\begin{figure}[b]
%\noindent\centering{
\begin{center}
\includegraphics[width=.50\textwidth]{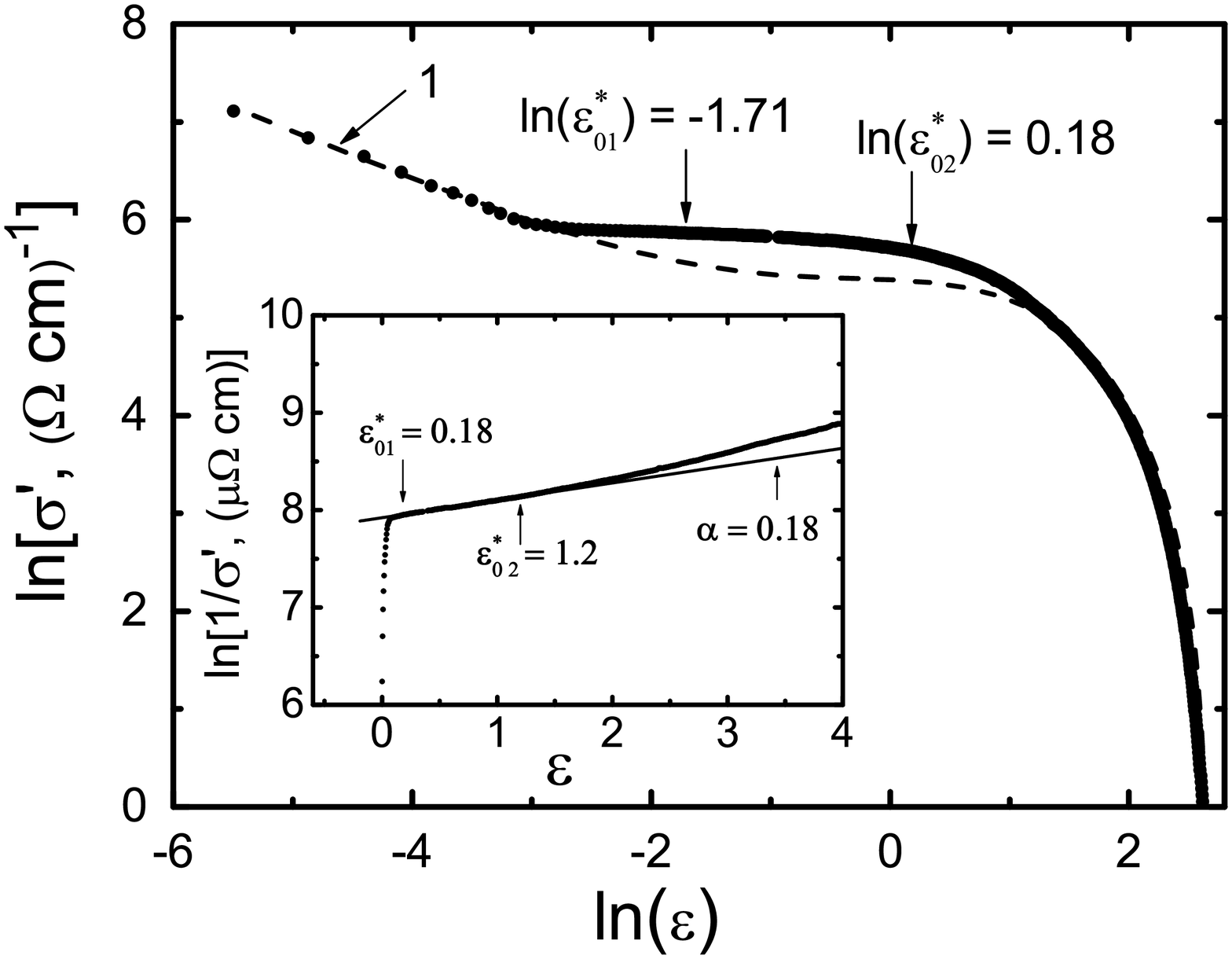}
\caption{$ln\sigma'$ vs $(ln\varepsilon)$ (dots) plotted over the whole temperature interval $T^*=171~K$ to $T_c^{mf}=11.2~K$ compared with Eq. (11) (dashed curve 1).
Insert: $ln\sigma'^{-1}$ (dots) as a function of $\varepsilon$. Solid line shows the linear extrapolation with the slope $\alpha=0.18$ which determines $\varepsilon_{c0}^* =1/\alpha = 5.6$  (see the text).}
\end{center}
%\label{ResTr2}
\end{figure}

\indent Above $T_0$ measured $\sigma'(\epsilon)$ noticeably upturns from the linear 3D AL dependence (Fig. 3, dots).
Such $\sigma'$ vs $\epsilon$ behavior is usually attributed to the MT \cite{Mak,Tho} fluctuation contribution to the excess conductivity in the 2D fluctuation region \cite{HL}.
Below $T_0$ near $T_c$, where $\xi_c(T) > d$, the fluctuating pairs have to interact in the whole sample volume, thus forming the 3D state.
But above $T_0$, where $\xi_c(T) < d$, the Josephson interaction  between the pairs in the whole sample volume is lost, suggesting a transition into 2D state \cite{Xie}.
Nevertheless, up to $T_{01}>T_0$ the Josephson interaction between the internal planes is believed to hold out \cite{S2,Xie}, and $\sigma'(\varepsilon)$ can be described by the MT term (5) of the Hikami-Larkin (HL) theory \cite{HL}.
It is believed that $\xi_c(T)=d$ at $T_0$ enabling the calculation of $\xi_c(0)$ as mentioned above.
Thus, $T_0$ is considered to be a crossover temperature corresponding to the 3D-2D and simultaneously to the AL-MT transition \cite{S2,S4,Xie}.

Finally, above $T_{01}$ (correspondingly above $\varepsilon_{01}$), at which
$\xi_c(T)=d_{01}$, the pairs are believed to be confined within As-Fe-As layers, or within $CuO_2$ planes in the case of cuprates, thus forming the quasi-2D conductivity \cite{Xie}.
Apparently, there is no direct interaction even between the internal planes now.
Here $d_{01} \ll d$ is the distance between As atoms in the conducting As-Fe-As layer or between the conductive $CuO_2$ planes, as in the cuprates.

As expected, above $T_0$ and up to $T_{01}\approx 42$\,K ($\ln \varepsilon_{01} \approx 1.0$) $\sigma'(T)$ is fitted by the MT fluctuation term (5) (Fig. 3, dashed curve 2) of the HL theory \cite{HL}
\begin{equation}
\sigma_{MT} '=\frac{e^2}{8\,d\,\hbar}\frac{1}{1-\alpha/\delta}\,ln\left((\delta/\alpha)\,
\frac{1+\alpha+\sqrt{1+2\,\alpha}}{1+\delta+\sqrt{1+2\,\delta}}\right)\,\varepsilon^{-1},
\label{MT}
\end{equation}
which dominates well above $T_c$ in the 2D fluctuation region \cite{HL,Xie,S4}.
In Eq. (5)
\begin{equation}
\alpha = 2\biggl[\frac{\xi_c(0)}{d}\biggr]^2\,\varepsilon^{-1}
\label{alp}
\end{equation}
is a coupling parameter,

\begin{equation}
\delta=\beta
\frac{16}{\pi\,\hbar}\biggl[\frac{\xi_c(0)}{d}\biggr]^2\,k_B\,T\,\tau_{\phi}
\label{TM}
\end{equation}
is the pair-breaking parameter,
$\tau_{\phi}$
that is defined by equation
\begin{equation}
\tau_{\phi}\beta\,T={\pi\hbar}/{8k_B\varepsilon}=A/\varepsilon
\label{tau}
\end{equation}
is the phase relaxation time, and $A=2.998\cdot 10^{-12}$ sK.
The factor $\beta=1.203(l\,/\,\xi_{ab}$), where $l$ is the mean-free path and $\xi_{ab}$ is the coherence length in the {\it ab} plane, considering the clean limit approach ($l>\xi$) \cite{S2,S4}.

Strictly speaking, our fit in the 2D fluctuation region (Fig. 3, curve 2) is not perfect.
Most likely it is due to the largely enhanced $\sigma'(T)$ above $T_0$ in comparison to that obtained for the YBCO films \cite{S2,S4}.
For the first time the enhancement of the excess conductivity in the 2D MT fluctuation region, marked in Fig. 3 as a maximal difference $\Delta (ln\sigma')$ between the data and extrapolated 3D AL term, was observed for $SmFeAsO_{0.85}$ \cite{S1}.
But now the enhancement, $\Delta (ln\sigma')$, is even larger (refer to Fig. 3).
The largest $\Delta(ln\sigma')$ was again observed for
the wholly magnetic superconductor $Dy_{0.6}Y_{0.4}Rh_{3.85}Ru_{0.15}B_4$ (Table I).
In this case, the $ln\sigma'$ vs $ln\varepsilon$ was found to be completely flat in the large temperature interval above $T_0$, being evidently behind any fluctuation description \cite{AT}.
In our case $ln\sigma'$ vs $ln\varepsilon$ is also somewhat close to be flat (Fig. 3, dots).
The result allows us to conclude that observed $\sigma'(T)$ increase above $T_0$ is likely due to expected enhanced magnetic interaction in studied $EuFeAsO_{0.85}F_{0.15}$.
Nevertheless we have applied Eq. (5) to fit the data.
Unfortunately neither $l$ \cite{S4} nor $\xi_{ab}$ \cite{Oh} are measured in our experiment, and $\tau_{\phi}$ remains uncertain.
That is why, we have used here somewhat another approach.
First, we set $\delta =2$, because in YBCO films it is always $\approx 2$ when $\xi_c(0)$ is properly defined \cite{S2,S4}.
Next, we have employed the following equality
\begin{equation}
\xi_c(0)= d\,\sqrt\varepsilon_0 = d_{01} \,\sqrt\varepsilon_{01},
\label{th}
\end{equation}
to rewrite Eq.(6) as
\begin{equation}
\alpha = 2\,\varepsilon_{01}/\varepsilon
%\label{\alpha}
\end{equation}
assuming $d=d_{01}$.
As one can see, only the value of $\varepsilon_{01}$ defines Eq. (5) now.
Here $\varepsilon_{01}$ corresponds to a temperature $T_{01}$ where the theoretical MT curve (Eq. (5)) finally deviates from or traverses the experimental data (Fig. 3, curve 2).
It is assumed that $\xi_c(T)$ which increases along with T decrease becomes equal to $d_{01}$ at $T_{01}$ \cite{S1} connecting the internal layers by the Josephson interaction \cite{Xie}.
As a result, just below $T_{01}$ somewhat correlated 2D fluctuation conductivity has to appear in the FeAs-base superconductor, as mentioned above.
It is important to note that, in accordance with the theory \cite{EK}, the stiffness of the wave function of the high-Tc superconductor persists just up to the temperature $T_{01}$  \cite{KMO}.
Thus, there is a definite connection between the crystal structure and physical properties of the HTS's emphasized by the extremely short coherence length (e.g $\xi_c(0)\approx 1.65\AA$ in optimally doped (OD) YBCO \cite{S2,S4}).
\begin{table}[tbp]
\caption [ ]

The parameters of the $YB_2Cu_3O_{7-\delta}$ (1),  $SmFeAsO_{0.85}$ (2), $EuFeAsO_{0.85}F_{0.15}$ (3) and $Dy_{0.6}Y_{0.4}Rh_{3.85}Ru_{0.15}B_{4}$ (4).
\centering
\begin{tabular}{||l|c|c|c|c|c|c|c||}
\hline\hline
Sample & $T_c$ & $T_c^{mf}$ & $T^*$ & $\Delta\,T_{3D}$ & $\xi_c(0)$ & $\Delta\,(ln\sigma')$ & 2$\Delta^*(T_c)/k_BT_c$\\
 & $(K)$& $(K)$ & $(K)$ & $(K)$ & $(\AA)$ &  & \\ [0.5ex]
\hline
1 & 87.4 & 88.5 & 203 & 1.8 & 1.65 & 0.2 & 5\\ [0.5ex]
\hline
2 & 55 & 57 & 175 & 1.5 & 1.4 & 0.5 & $\sim 5$\\ [0.5ex]
\hline
3 & 11 & 11.2 & 171 & 0.5 & 2.84 & 1.4 & 4.4\\
\hline
4 & 6.4 & 6.68 & 161 & 0.16 & 2.9 & 1.7 & - \\
\hline\hline
\end{tabular}
\label{tab:sample-values}
\end{table}

\indent This approach has provided a very good 2D MT fit in the case of $SmFeAsO_{0.85}$ \cite {S1}.
Importantly, it was found that $d_{01} \approx 3\AA$ in this case, which is about the width of the As-Fe-As layer $d_{01} =2.75\AA$
reported in the literature for $SmFeAsO$ \cite{Karp}.
In the case of $EuFeAsO_{0.85}F_{0.15}$, $ln\varepsilon_{01} \approx 1.0$ (Fig. 3, curve 2) resulting in $\varepsilon_{01}= 2.72$ which we used while computing Eq's. (5 and 10).
It seems to be the most self-consistent result of our 2D fluctuation treatment.
However, from Eq.(9) $d_{01}$ appears to be only $\sim 1.72\AA$ in this case.
We could not find information about the width of the As-Fe-As layer in
$EuFeAsO_{0.85}F_{0.15}$, but one may assume that it can differ from $d_{01} =2.75\AA$
a little bit.
On the other hand, if we use the common approach and take the coupling parameter $\alpha$ from Eq. (6) with d=c=13\AA\,, it will result in dashed curve (3) (Fig. 3), which is close to that found for YBCO film  \cite {S2} but evidently does not meet the case.
Eventually, all these considerations allow us to conclude that the role of magnetic interaction in studied $EuFeAsO_{0.85}F_{0.15}$ is expectedly larger than even in $SmFeAsO_{0.85}$.
However, all these discrepancies as for 2D MT fluctuation contribution do not affect our further analysis.
Really, to proceed with the analysis we need only the value of $\xi_c(0)$ which is strictly determined by the crossover temperature $T_0$.\\

\indent {\bf 3.2.\, Pseudogap analysis}\\

Evidently, to get any information about PG from the excess conductivity one needs an equation which describes the whole experimental curve, from $T^*$ down to $T_c$, and contains the parameter $\Delta^*$ in the explicit form.
In cuprates $\Delta^*$ is referred to as a pseudogap parameter which is most likely due to the local pair formation and has to reflect the peculiarities of the LP interaction along with decrease of temperature from $T^*$ down to $T_c$ \cite{S2,S6,PVB}.
In $EuFeAsO_{0.85}F_{0.15}$ $\Delta^*$ is assumed to be due to both local pairs and magnetic interaction, as mentioned above.
Thus, its temperature dependence is expected to somehow reflect the complex interplay between the superconducting fluctuations and magnetism which is of a primarily importance to comprehend the principles of the coupling mechanism in HTS's.
\begin{figure}[b]
%\noindent\centering{
\begin{center}
\includegraphics[width=.50\textwidth]{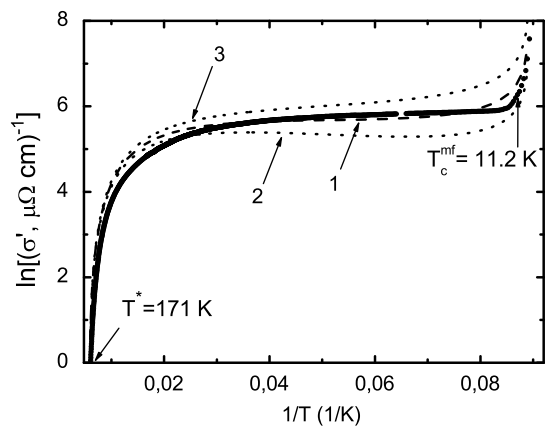}
\caption{$ln\sigma'$ as a function of 1/T (dots) in comparison with Eq. (11) plotted with different $\Delta^*(T_c)/k_BT_c$. The best fit is obtained for $\Delta^*(T_c)/k_BT_c=2.2$ (dashed curve 1). Also shown are the results of fitting with $\Delta^*(T_c)/k_BT_c =2.8$ (dotted curve 2) and $\Delta^*(T_c)/k_BT_c =1.6$ (dotted curve 3).}
\end{center}
%\label{ResTr2}
\end{figure}

Because of absence of the complete fundamental theory, we have applied the LP model approach to perform the PG analysis.
The equation for $\sigma'(\varepsilon)$ has been proposed in Ref. \cite{S6} with respect to the local pairs
\begin{equation}
\sigma '(\varepsilon) = \frac{e^2\,A_4\,\left(1 -
\frac{T}{T^*}\right)\,\left(exp\left(-\frac{\Delta^*}{T}\right)\right)}{(16\,\hbar\,\xi_c(0)\,
\sqrt{2\,\varepsilon_{c0}^*\,\sinh(2\,\varepsilon\,/\,\varepsilon_{c0}^*})}.
\label{sigma-eps}
\end{equation}
Here, the dynamics of pair-creation $(1-T/T^*)$ and pair-breaking $exp(-\Delta^*/T)$
below $T^*$ has been taken into account in order to correctly describe the experiment \cite{S2,S6}.
Solving Eq. (11) regarding $\Delta^*(T)$ one can readily obtain
\begin{equation}
\Delta^*(T) = T\,ln\frac{e^2\,A_4\,(1 - \frac{T}{T^*})}{\sigma '(T)\,16\,\hbar\,\xi_c(0)\,\sqrt{2\,\varepsilon_{c0}^*\,\sinh(2\,\varepsilon\,/\,\varepsilon_{c0}^*)}},
\label{delta-t}
\end{equation}
where $A_4$ is a scaling factor which has the same meaning as the C-factor in the FLC theory \cite{S2,Oh,S6} and $\sigma'(T)$ is the experimentally measured excess conductivity over the whole temperature interval from $T^*$ down to $T_c^{mf}$.
In the case of YBCO films \cite{S2,S6} and BiSrCaCuO single crystals \cite{Tk} Eq. (11) fits the data extremely good,
thus demonstrating the validity of this description.
From our point of view it also means that found by means of Eq. (12) $\Delta^*(T)$ has to properly reflect the properties of the pseudogap \cite{S2,S6}.

The next step of the LP model approach is to determine some additional unknown parameters important for the further analysis.
Apart from $T^*$, $T_c^{mf}$ and $\xi_c(0)$ determined above, both Eq. (11) and Eq. (12) contain the theoretical parameter $\varepsilon^*_{c0}$, numerical factor $A_4$, and $\Delta^*(T_{c})$, which is the PG value at $T_c^{mf}$.
Nevertheless, all the parameters can directly be determined from the experiment.
First, in the range of $ln\varepsilon^*_{01}<ln\varepsilon^*<ln\varepsilon^*_{02}$ (Fig. 4)  or accordingly $\varepsilon^*_{01}<\varepsilon^*<\varepsilon^*_{02}$ ($13.2~K<T<24.6~K$) (see insert in Fig. 4),
$\sigma'^{-1} \sim$ exp$(\varepsilon$).
This exponential dependence turned out to be the common feature of many HTS's \cite{S2,S6,B}.
As a result, ln($\sigma'^{-1}$) is a linear function of $\varepsilon$ with a slope $\alpha^*$=0.18 which determines parameter $\varepsilon_{c0}^* = 1/\alpha^*$=5.54  \cite{S6,B}.

To find $A_4$ we calculate $ln\sigma'(ln\varepsilon)$ using Eq. (11) in the whole temperature interval from $T^*$ and down to $T_c$, but fit experiment in the range of 3D AL fluctuations near $T_c$ (Fig. 4), where $ln\sigma'(ln\varepsilon)$ is a linear function of the reduced temperature $\varepsilon$ with a slope $\lambda$ = $-$1/2 \cite{S2,S6}.
In contrast to YBCO films \cite{S2,S6} and BiSrCaCuO single crystals \cite{Tk} the curve given by Eq.(11) noticeably deviates down from the data above $T_0$ now (Fig.~4).
The deviation is most likely the result of enhanced magnetic interaction, as mentioned above.
But it is of no importance for the further consideration since the value of $A_4$ can be strictly determined from the plot.
As it is seen from the figure, the fit in the range of the 3D AL fluctuations near $T_c$ is expectedly good resulting in $A_4 = 2.8$.
Importantly, if we put found rather unusual $\Delta^*(T)$ (Fig.~6) into Eq. (11) instead of constant $\Delta^*(T_{c})$,
the resulting curve will describe the $\sigma'(T)$ data perfectly.

Next, in our consideration $\Delta^*(T_{c}) = \Delta(0)$ is assumed, where $\Delta(0)$ is the superconducting gap at T=0 \cite{Sta,Yam}.
Thus, the equality 2$\Delta^*(T_{c})/k_BT_c$ = 2$\Delta(0)/k_BT_c$ is to occur.
Finally, to estimate $\Delta^*(T_{c})$, which we use in Eq. (11) to initiate the analysis, we plot $ln\sigma'$ as a function of 1/T (Fig.~5, dots) \cite{S2,PVB} and fit it by Eq. (11).
In this case the slope of the theoretical curve (Fig. 5, dashed and dotted curves 1-3) turns out to be very sensitive to the value of $\Delta^*(T_{c})$ \cite{S2,S6}.
%Again, the fit is not completely good most likely due to influence of \textcolor{red}{the} magnetism.
The best fit is obtained when 2$\Delta^*(T_{c})/k_B\,T_c\approx 4.4$ (Fig. 5. dashed curve 1) which is close to the BCS value for the superconductors with strong coupling \cite{Jee1,AK}.
The result suggests that $\Delta^*(T_{c})/k_B\approx 24.2\,K$ ($\approx 2.1 \,meV$).
It seems to be reasonable seeing that measured $T_c =11\,K$ is relatively low.
Thus, all parameters needed to calculate $\Delta^*(T)$ are determined now.
Fig.~6 (dots) displays found $\Delta^*(T)$ calculated using Eq. (12) with the following set of parameters derived from the experiment:
$T^* = 171$\,K, $T_c^{mf} = 11.2$\,K, $\xi_c(0) = 2.84$\AA, $\varepsilon_{c0}^* = 5.6$,
$A_4 = 2.8$.

\begin{figure}[b]
%\noindent\centering{
\begin{center}
\includegraphics[width=.50\textwidth]{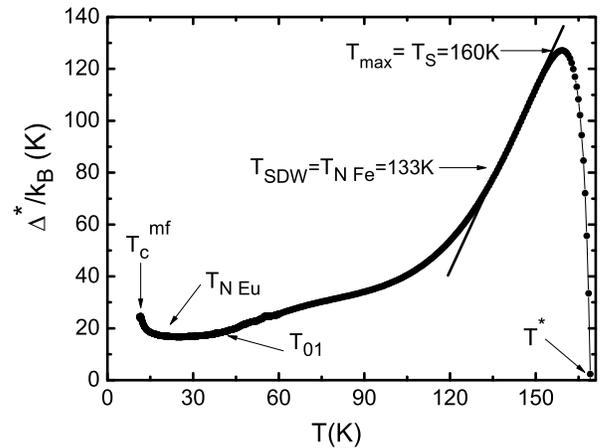}
\caption{$\Delta^*/k_B$ (dots) as a function of T. Solid line designates the positive slope linear region between $T_S=160\,K$ and
$T_{SDW}=T_{N Fe}=133\,K$,
which is believed to be the direct sign of the magnetic interaction in the HTS's.
Also shown are the representative temperatures $T^* =171~K$, $T_{01} =42~K$, $T_{NEu}=20~K$, and $T_c^{mf}=11.2~K$.}
\end{center}
%\label{ResTr2}
\end{figure}

As can be seen from the figure, $\Delta^*(T)$ exhibits narrow maximum at $T_{max} \approx 160$\,K followed by a positive slope linear region down to $T\approx 133\,K$ (Fig. 6, dots).
The shape of the whole curve is completely different from that usually observed for YBCO and BiSCCO cuprates, where $\Delta^*$ is an increasing function of temperature with the wide maximum at $T_{pair} \approx 130\,K$ and $\approx 150\,K$, respectively \cite{S2,Tk}.
However, the curve turns out to be typical for the "1111" FeAs-based superconductors.
For the first time such positive slop linear $\Delta^*(T)$ dependence was observed for $SmFeAsO_{0.85}$ between $T_s = 150\,K$ and $T_{SDW} = 133\,K$, and is believed to be the noticeable feature of the magnetic influence in the high-$T_c$ superconductors \cite{S1}.
In $SmFeAsO$ both representative temperatures $T_s =150\,K$ and $T_{SDW}= T_{NFe}= 133\,K$ were independently determined from the resistivity \cite{Sad,Ren2} and specific heat \cite{Ding} experiments, respectively.
By analogy with that results, we may conclude that $T_{max}=T_s \approx 160\,K$ is the structural transition temperature, and the next representative temperature is $T_{SDW}=T_{NFe}\approx 133\,K$, which corresponds to the SDW transition followed most likely by the AF ordering of Fe spins in $EuFeAsO_{0.85}F_{0.15}$.
Found $T_s \approx 160\,K$ is higher than that observed for $SmFeAsO$ \cite{Ren} and $LaFeAsOF$ \cite{Kam,Sad}.
It is likely because the Eu-based compounds (e.g. $EuFe_2As_2$) demonstrate the highest $T_S$ \cite{Jee1,Ren}, as mentioned above.
But the second representative temperature in $EuFeAsO_{0.85}F_{0.15}$, namely  $T_{SDW}=T_{NFe}\approx 133\,K$, is just the same as found for the $SmFeAsO$, and was distinctly observed for the first time.
Below this temperature $\Delta^*(T)$ continues to decrease gradually down to $T\approx 30\,K$.
Then it starts to increase with the more pronounced rise just below $T_{NEu}\approx 20\,K$, which is the temperature of Eu 4f moments ordering \cite{Jee1,Ren}.
And finally $\Delta^*(T)/k_B$ acquires the value of about 24\,K at $T=T_c^{mf}$ in a good agreement with the above calculations.
Note, that $\Delta^*(T)$ changes its curvature, from negative to the positive one, just around $T_{01}=42\,K$ (Fig. 6), below which the fluctuation conductivity has to appear.
Thus, we may conclude that, despite of the strong influence of magnetism, our LP model approach has allowed us to obtain
rather reasonable and self-consistent results.
\begin{figure}[b]
%\noindent\centering{
\begin{center}
\includegraphics[width=.50\textwidth]{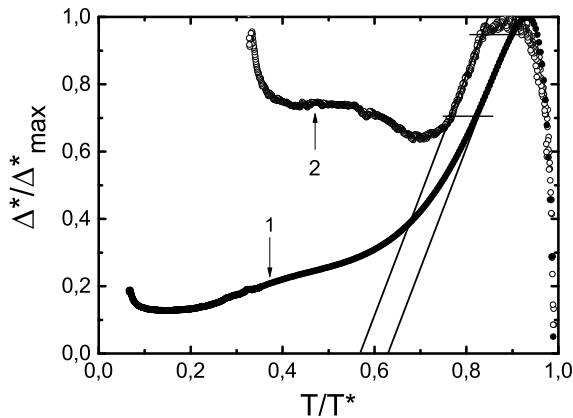}
\caption{$\Delta^*(T)/\Delta_{max}$ as a function of $T/T^*$ for studied $EuFeAsO_{1-x}F_x$ (curve 1, dots) and reference $SmFeAsO_{0.85}$ \cite{S1} (curve 2, circles). Solid lines with equal positive slope designate the linear $\Delta^*(T)$ regions for both samples. Horizontal lines designate its equal length. The result suggests the generality of the interaction mechanism for the superconductors in which the AF ordering may coexist with superconductivity.}
\end{center}
%\label{ResTr2}
\end{figure}

To be more sure we have compared results (Fig. 7 curve 1) with those obtained for $SmFeAsO_{0.85}$ (Fig. 7, curve 2).
The result of the comparison is plotted in Fig. 7 in double reduced scale.
In the case of $SmFeAsO_{0.85}$ \cite{S1} the positive slope linear drop of $\Delta^*(T)$ was qualitatively explained within the Machida-Nokura-Matsubara (MNM) theory developed for the superconductors in which the AF ordering may coexist with the superconductivity, such as for example $RMo_6S_8$ (R = Gd, Tb, and Dy) \cite{MNM}.
In accordance with the MNM theory, in such compounds $\Delta(T)$ linearly drops below $T_N <T_c$
due to formation of the energy gap of SDW on the Fermi surface which partially suppresses the SC gap.
As AF gap saturates at lower temperatures, $\Delta(T)$ gradually recovers its value with increasing the SC condensation energy.
The observation of the similar $\Delta^*(T)$ behavior in $SmFeAsO_{0.85}$ but above $T_c$ (Fig. 7, curve 2) was considered to be an additional evidence for the LP existence in the FeAs-based superconductors \cite{S1,S2}.
Really, it was assumed that, in accordance with the MNM theory, the order parameter of the local pairs, $\Delta^*$, is suppressed below $T_s$ by the low-energy magnetic fluctuations resulting in observed positive slope linear drop of $\Delta^*(T)$ followed by the SDW transition \cite{S1,Dr,San,Che3,Zhu,Fer}.
As it is seen in Fig. 7, both samples demonstrate just the same positive slope linear drop of $\Delta^*(T)$ just between $T_s$ and $T_{SDW}$, which is designated by the straight lines in the figure.
Moreover, the length of the both positive slope regions turned out to be also the same suggesting the same mechanism of the magnetic interaction in both superconductors.
However, in contrast with $SmFeAsO_{0.85}$, in $EuFeAsO_{0.85}F_{0.15}$ $\Delta^*(T)$ continues to fall even below $T_{SDW}$ pointing out the more strong influence of magnetism in this case.
Summarizing, we may conclude that obtained for $EuFeAsO_{0.85}F_{0.15}$ $\Delta^*(T)$
also can be qualitatively explained within the MNM theory but likely with the larger
value of the SDW gap.
Thus, in contrast with results of Ref.\cite{25,Jon}, we may conclude that influence of the electron scattering due to $Eu^{2+}$ local moments also must be take into account to explain revealed $\Delta^*(T)$ in $EuFeAsO_{0.85}F_{0.15}$. \\

\indent {\bf 4.\, CONCLUSION}\\

For the first time the excess conductivity $\sigma'(T)$ in the FeAs-based superconductor $EuFeAsO_{0.85}F_{0.15}$ was analysed over the whole temperature range $T^* = 171\,K$ to $T_c^{mf} =11.2\,K$.
$EuFeAsO_{0.85}F_{0.15}$ is characterized by the enhanced magnetic interaction mostly
owing to the large additional magnetic moment of about $7\mu_B$ carried by Eu \cite{Jee1,Ren}.
Nevertheless, the analysis was performed within the LP model developed for cuprate HTS's \cite{S2,S6} and based on the assumption of the local pair formation below $T^*\gg T_c$ \cite{L,H,Eng}.
In magnetic superconductor such as $EuFeAsO_{0.85}F_{0.15}$ the temperature dependence of $\sigma'$ has to reflect the complex interplay between superconducting fluctuations and magnetism which is of a primarily importance to comprehend the principles of the coupling mechanism in HTS's.
Naturally, we expected to reveal the $\sigma '(T)$ peculiarities caused by the magnetism.

It was shown that over the temperature range $T_c^{mf}$ to $T_0$ = 11.7~K ($\ln \varepsilon_0 \approx -3.04$) $\sigma'$ vs $\epsilon$ is expectedly fitted by the 3D fluctuation term (3) of the AL theory (refer to Fig. 3, solid line 1).
Above $T_0$ up to $T_{01}\approx 42$\,K ($\ln \varepsilon_{01} \approx 1.0$) $\sigma'(T)$ can be described by the 2D MT fluctuation term (5) (Fig. 3, solid curve 2) of the Hikami-Larkin theory \cite{HL}.
It means that conventional fluctuating Cooper pairs have to be present in the superconductor near $T_c$, and the stiffness of the SC wave function \cite{EK} persists at least up to $T_{01}$ \cite{KMO}.
However, the range of the 3D AL fluctuations $\Delta T_{3D}=0.5\,K$ is relatively short (Fig. 2 and 3) in comparison with
the YBCO films \cite{S2} and $SmFeAsO_{0.85}$ polycrystals \cite{S1}, and enhanced $\sigma'(T)$ is observed above $T_0$.
The shortest $\Delta T_{3D} \approx 0.16\, $ and the largest $\sigma'(T)$ enhancement in the 2D fluctuation region were observed for the wholly magnetic superconductor
$Dy_{0.6}Y_{0.4}Rh_{3.85}Ru_{0.15}B_4$ with $T_c = 6.4\,K$ \cite{AT} (see the Table).
Thus, the shortening of the $\Delta T_{3D}$ and enhanced $\sigma'_{2D}$
can be considered as an evidence of the enhanced magnetic interaction in $EuFeAsO_{0.85}F_{0.15}$.
Nevertheless, the strictly designated in the experiment $T_0$ allows us to determine $\xi_c(0)= (2.84 \pm 0.02)\AA$
being of the essential importance for the further analysis.

Making use of Eq. (11) and Eq. (12) to analyze the $\sigma'(T)$, temperature dependence of the parameter $\Delta^*$ was calculated over the whole temperature range $T^*= 171\,K$ down to $T_c^{mf} = 11.2\,K$ (refer to Fig. 6).
In cuprates $\Delta^*$ is referred to as a pseudogap parameter which is most likely due
to the local pair  formation at $T<T^*$ and has to reflect the peculiarities of the LP
interaction along with decrease of temperature \cite{S2,S6,PVB}.
In $EuFeAsO_{0.85}F_{0.15}$ the more complicated character of revealed $\Delta^*(T)$ (Fig. 6, dots) suggests a conclusion that additional magnetic interaction also has to be taken into account.
Really, found $\Delta^*(T)$ exhibits narrow maximum at $T_s\approx 160$\,K followed by the positive slope linear region down to $T_{SDW}=T_{NFe}\approx 133\,K$ (Fig. 6, dots).
The shape of the whole curve is completely different from that usually observed for YBCO and BiSCCO cuprates  \cite{S2,Tk}.
For the first time such positive slope linear $\Delta^*(T)$ dependence was observed for $SmFeAsO_{0.85}$ between $T_s = 150\,K$ and $T_{SDW} = 133\,K$ and is believed to be the more noticeable feature of the magnetic influence in the high-$T_c$ superconductors \cite{S1,S2}.
Found $T_s \approx 160\,K$ is higher than that observed for $SmFeAsO$ \cite{Ren} and $LaFeAsOF$ \cite{Kam,Sad}.
It is likely because the Eu-based compounds (e.g. $EuFe_2As_2$) demonstrate the
highest $T_S$ \cite{Jee1,Ren}.
But the SDW temperature $T_{SDW}=T_{NFe}\approx 133\,K$ is the same as in $SmFeAsO$, and distinctly revealed for the first time.
Below this temperature $\Delta^*(T)$ continues decrease gradually down to $T\approx 30\,K$.
Then it starts to increase more rapidly just below $T_{NEu}\approx 20\,K$ which is the temperature of Eu 4f moments ordering \cite{Jee1,Ren}.
And finally $\Delta^*(T)/k_B$ acquires the value of about 24\,K at $T_c^{mf}$ in a good agreement with our calculations.
Thus, we may conclude that, despite the strong influence of magnetism, our LP model approach has allowed us to obtain rather reasonable and self-consistent results.
This experimental fact points out at the possibility of the local pair existence even in magnetic superconductors.

\indent To be more sure we have compared results (Fig. 7 curve 1) with those obtained for $SmFeAsO_{0.85}$ (Fig. 7, curve 2).
Importantly, both samples demonstrate just the same positive slope linear drop of $\Delta^*(T)$ just between $T_s$ and $T_{SDW}$, which is designated by the straight lines in the figure.
Moreover, the length of the positive slope regions turned out to be also the same suggesting the same mechanism of the magnetic interaction in both superconductors.
However, in $EuFeAsO_{0.85}F_{0.15}$ $\Delta^*(T)$ continues to fall even below $T_{SDW}$ pointing out the more strong influence of magnetism in this case.
In $SmFeAsO_{0.85}$ \cite{S1} such unusual $\Delta^*(T)$ behavior was qualitatively explained within the MNM theory \cite{MNM} in which the $\Delta^*(T)$ drop was assumed to be due to formation of the energy gap of SDW on the Fermi surface which partially suppresses the SC gap.
The observation of the linear $\Delta^*(T)$ behavior in $SmFeAsO_{0.85}$ above $T_C$ (Fig. 7, curve 2) was considered as an additional evidence for the LP existence in the FeAs-based superconductors \cite{S1,S2}.
It was assumed that, in accordance with the MNM theory, the order parameter of the local pairs, $\Delta^*$, is suppressed below $T_s$ by the low-energy magnetic fluctuations resulting in observed linear drop of $\Delta^*(T)$ followed by the SDW transition \cite{S1,Dr,San,Che3,Zhu,Fer}.
By analogy we may conclude that in $EuFeAsO_{0.85}F_{0.15}$ the local pairs also have to be taken into account at $T<T^*$, and found $\Delta^*(T)$
also can be qualitatively explained within the MNM theory but likely with the larger
value of the SDW gap.

Recently the similar $\Delta^*(T)$ dependence was observed in YBCO-PrBCO superlattices and YBCP-PrBCO sandwiches \cite{SR} as well as in $HoBa_2Cu_3O_{7-\delta}$ slightly doped single crystals \cite{S3}  suggesting the generality of the interaction mechanism for the superconductors in which the AF ordering may coexist with superconductivity.\\

\indent {\bf \, Acknowledgement}\\

The research was supported by WRC EIT+ within the project "The
Application of Nanotechnology in Advanced Materials" – NanoMat
(POIG.01.01.02-02-002/08) financed by the European Regional Development
Fund (IEOP 1.1.2).

\end{document}